\documentclass{aa}  

\usepackage{graphicx}
\usepackage{txfonts}
\usepackage{caption}
\usepackage{subcaption}

\usepackage[export]{adjustbox}

\begin{document}

  \title{Rossby waves on stellar equatorial $\beta$- planes: uniformly rotating radiative stars}

   \subtitle{}

   \author{M. Albekioni\inst{1,2,3}, T.V. Zaqarashvili\inst{4,2,3} and V. Kukhianidze\inst{2,3}
          }

   \institute{
   Georg-August-Universit\"at, Institut f\"ur Astrophysik, Friedrich-Hund-Platz 1, 37077, G\"ottingen, Germany\\
   \and
   Department of Astronomy and Astrophysics at Space Research Center, School of Natural Sciences and Medicine, Ilia State University, Kakutsa Cholokashvili Ave. 3/5, Tbilisi 0162, Georgia\\
   \and
   Evgeni Kharadze Georgian National Observatory, Abastumani, Adigeni 0301, Georgia\\
   \and
   Institut of Physics, IGAM, University of Graz, Universit\"atsplatz 5, 8010 Graz, Austria\\
             }

   \date{}
 
  \abstract
   {Rossby waves arise due to the conservation of total vorticity in rotating fluids and may govern the large-scale dynamics of stellar interiors. Recent space missions collected huge information about the light curves and activity of many stars, which triggered observations of Rossby waves in stellar surface and interiors. }
   {We aim to study the theoretical properties of Rossby waves in stratified interiors of uniformly rotating radiative stars with sub-adiabatic vertical temperature gradient. }
   {We use the equatorial beta-plane approximation and linear vertical gradient of temperature to study the linear dynamics of equatorially trapped Rossby and inertia-gravity waves in interiors of radiative stars. The governing equation is solved by the method of separation of variables in the vertical and latitudinal directions.   }
   {Vertical and latitudinal solutions of the waves are found to be governed by Bessel functions and Hermite polynomials, respectively. Appropriate boundary conditions at stellar surface and poles define analytical dispersion relations for Rossby, Rossby-gravity and inertia-gravity waves. The waves are confined in surface layers of 30-50 $H_0$, where $H_0$ is surface density scale height, and trapped between the latitudes of $\pm 60^0$. Observable frequencies (normalized by the angular frequency of the stellar rotation) of Rossby waves with $m=1$ ($m=2$), where $m$ is the toroidal wavenumber, are in the interval of 0.65-1 (1.4-2), depending on stellar rotation, radius and surface temperature.}
   {Rossby-type waves can be systematically observed using light curves of Kepler and TESS stars. Observations and theory then can be used for the sounding of stellar interiors.}

\titlerunning{.}
\authorrunning{Albekioni et al.}

   \maketitle

\section{Introduction}

Rossby (planetary) waves are essential features of large-scale dynamics of rotating fluids. The theoretical background for the waves started by \citet{Hadley1735}, while Laplace tidal equations created the basic for the mathematical description \citep{Laplace1893}. \citet{Hough1897,Hough1898} found the solutions of the Laplace equations for the Rossby waves (low frequency solution in Hough's terminology). However, the physical meaning of the waves was first described by  \citet{Rossby1939}: the waves arise due to the conservation of total vorticity (planetary+relative) on a rotating sphere, which drives the propagating oscillations in the opposite direction of rotation. 

Rossby waves are well studied in the Earth's context by observations and theory. Westward propagating waves with predicted phase speed have been continuously observed in the terrestrial atmosphere \citep{Hovmoller1949, Eliasen1965, Yanai1983, Lindzen1984, Hirooka1989, Madden2007} and oceans \citep{Chelton1996, Hill2000}. The theory of terrestrial Rossby waves is also well studied \citep{Haurwitz1940, Lindzen1967, Gill1982,Pedlosky1987}. The detailed dynamics of the Rossby waves in the Earth's atmosphere and oceans is summarized in reviews of \citet{Platzman1968} and \citet{Salby1984}. 

Recent direct observations of Rossby waves in the solar surface \citep{Lopten2018} revived the interest for the study of Rossby waves on the Sun. The waves were observed by different  methodologies such as granular tracking and helioseismology \citep{Liang2019,Hanasoge2019,Proxauf2020,Gizon2021,Hanson2022}. The Rossby wave signature was also found in the dynamics of solar coronal bright points \citep{McIntosh2017, Krista2017}. It was suggested that the magnetic Rossby waves may influence the short-term activity variations in the solar dynamo layer below the convection zone \citep{Zaqarashvili2010, Zaqarashvili2018, Dikpati2018, Dikpati2020}. The Rossby waves are important in the dynamics of many astrophysical objects such as solar system planets, exoplanets, accretion disks etc. \citep{Zaqarashvili2021}.

Recent space missions CoRoT, Kepler and TESS collected huge information about stellar light curves and activity. \citet{VanReeth2016} reported the detection of Rossby waves in rapidly rotating $\gamma$ Dor stars in period spacing patterns. The pressure field of Rossby waves on the stellar surface may influence their light curves as suggested by \citet{Saio2018}, hence the waves have been continuously observed in many Kepler stars with different spectral types \citep{Saio2018,Li2019,Jeffery2020,Samadi2020, Takata2020,Henneco2021,Saio2022}. \citet{Lanza2019} showed that the Rossby waves may represent a source of confusion in the case of slowly rotating inactive stars that are preferential targets for radial velocity planet search. Therefore, the theoretical description of stellar Rossby waves is important for stellar activity and exoplanetary research.

Rossby waves are known as r-modes in stellar society since \citet{Papaloizou1978}. \citet{Provost1981} made significant progress in the theoretical study of r-modes 
in uniformly rotating stars by perturbation analysis (see also \citet{Damiani2020} for the similar consideration). \citet{Saio1982} examined the r-mode oscillations in massive zero-age main sequence and ZZ Ceti stars. The studies generally concerned the slowly rotating stars, which allow the perturbation analysis by small expansion parameter for the slow rotation (note that \citet{Papaloizou1978} studied the rapidly rotating stars, but for high order harmonics of r-modes). On the other hand, it is of vital importance to study the Rossby wave for stars with any rotation rate. 

Here we aim to study the Rossby waves in stratified stellar interiors without approximation of slow rotation. We will use the formalism of terrestrial Rossby waves, which is well tested in the Earth atmosphere (e.g. \citet{Lindzen1967}). The formalism allows us to derive the exact solutions and dispersion relations for Rossby, Rossby-gravity and inertia-gravity waves in rectangular equatorial beta-plane approximation. The periods of different harmonics can be used for observations of the waves in stellar light curves. In this paper, we concern the radiative stars without an outer convection zone.

\section{Governing Equations}

We start with linearised momentum, continuity and energy equations in a frame of uniformly rotating star: 
 \begin{equation}
     {\rho_0}\frac{\partial \vec{v}}{\partial t} + 2 {\rho_0}{\vec\Omega}\times  \vec{v} = - \nabla p' + \rho' \vec {g},
     \label{Eq.(1)}
   \end{equation}
   \begin{equation}
     \frac{\partial \rho'}{\partial t} + (\vec{v}\cdot \nabla)\rho_0 + \rho_0 \nabla \cdot \vec{v} =0,
     \label{Eq.(2)}
     \end{equation}
     \begin{equation}
     \frac{\partial p'}{\partial t} + (\vec{v} \cdot \nabla) p_0 + \gamma p_0 \nabla \cdot \vec{v} =0,
     \label{Eq.(3)}
     \end{equation}
     where $\vec{v}$ is the velocity,  $\rho_0$ ($p_0$) is the unperturbed density (pressure),  $p'$ ($\rho'$) is the perturbation of pressure (density), $\vec{g}$ is the gravitational acceleration, $\vec \Omega$ is the angular velocity of rotation and $\gamma=c_p/c_v$ is the ratio of specific heats. 

In the following, we adopt the Cartesian coordinates $(x, y, z)$, where $x$ is directed towards rotation, $y$ is directed towards the north pole and $z$ is directed vertically upwards. Undisturbed medium is assumed to be in vertical hydrostatic balance  
\begin{equation}
\frac{d p_0}{d z}=-g \rho_0
\label{Eq.(4)}
\end{equation}
and the ideal gas law is written as   
\begin{equation}
p_0=\frac{k_b}{m}\rho_0 T=g\rho_0 H,
\label{Eq.(5)}
\end{equation}
where $k_b$ is the Boltzman constant, $T(z)$ is the temperature, $m$ - is the mass of hydrogen atom,  $H(z)=k_b T(z)/m g$ is the density scale height. Then the substitution of Eq.~(\ref{Eq.(5)}) into Eq.~(\ref{Eq.(4)}) gives the vertical distribution of the density governed by the equation
	\begin{equation}
	\frac{d \rho_0}{d z} = - \frac{\rho_0}{H} \left(1+ \frac{d H}{d z} \right).
	\label{Eq.(6)}
	\end{equation}

We closely follow to the formalism of \cite{Lindzen1967}, who considered the beta-plane approximation for the uniform temperature with depth. Our calculation, however, is performed for non-uniform distribution of temperature with depth.
Here we use the vertically hydrostatic assumption, which means that the vertical distribution of the pressure is only slightly disturbed from its static form as it is typical for geophysical/astrophysical flows with small Rossby number. This means that the vertical velocity is small and it is neglected in the vertical momentum equation, while it is kept in the continuity equation. This approximation avoids the consideration of internal gravity and acoustic waves and hence it is valid only for long-period waves. 

We change variables as 
	\begin{equation}
	{\tilde v_x}=\sqrt{\rho_0} v_x,  
	{\tilde v_y}=\sqrt{\rho_0} v_y,  
    {\tilde v_z}=\sqrt{\rho_0} v_z, 
    {\tilde p'}=\frac{p'}{\sqrt{\rho_0}} , {\tilde \rho'}=\frac{\rho'}{\sqrt{\rho_0}},
    \label{Eq.(7)}
	\end{equation}
so that Eqs.~(\ref{Eq.(1)}) - (\ref{Eq.(3)}) can be written after Fourier transform with $e^{i(- \sigma t + k x )}$ as
   	\begin{equation}
	-i \sigma v_x - f v_y = -i k p',
	\label{Eq.(8)}
	\end{equation}
	\begin{equation}
	-i \sigma v_y + f v_x = - \frac{\partial p'}{\partial y},
	\label{Eq.(9)}
	\end{equation}
	\begin{equation}
\frac{\partial p'}{\partial z} - \frac{1}{2 H} \left (1+\frac{d H}{d z}\right ) p'= - g \rho',
\label{Eq.(10)}
	\end{equation}
	\begin{equation}
	-i \sigma \rho' + i k v_x + \frac{\partial v_y}{\partial y} + \frac{\partial v_z }{\partial z}  - \frac{1}{2 H} \left( 1+\frac{d H}{dz} \right) v_z= 0,
	\label{Eq.(11)}
	\end{equation}
	\begin{equation}
	i \sigma p' = i\gamma g H \sigma \rho' - g\left [1-\gamma\left (1+\frac{d H}{dz} \right )\right ]v_z,
	\label{Eq.(12)}
	\end{equation}

where $f=2\Omega \sin \theta$ is the Coriolis parameter with $ \theta$ being a latitude (note that tilde sign is dropped from the variables). Due to the vertically hydrostatic assumption, the two terms, $-i\sigma v_z$ and $-2\Omega_y v_x$, were omitted on the left hand side of Eq.~ (\ref{Eq.(10)}). The ratios of  the omitted terms and the first term on right hand side of Eq.~ (\ref{Eq.(10)}) are proportional to $\sigma^2 H/g$ and $R H/\lambda^2$, respectively, where $R$ is the radius of the sphere and $\lambda$ is the horizontal wavelength. The first ratio is very small for the Rossby wave time scales. The second ratio is proportional to $H/R\ll 1$ for the typical Rossby wavelengths. Hence, the both terms are small and vertically hydrostatic assumption is justified in the current consideration of Rossby-type waves.

Eqs. ~(\ref{Eq.(8)}) and (\ref{Eq.(9)}) lead to the equation
\begin{equation}
	(f^2 - \sigma^2) v_y = i \sigma \left(\frac{k}{\sigma } f + \frac{\partial }{\partial y} \right) p'.
	\label{Eq.(13)}
	\end{equation}
Eliminating $\rho'$, $v_z$, $v_x$ from Eqs. ~(\ref{Eq.(10)})-(\ref{Eq.(12)}), we obtain
$$ \frac{\partial }{\partial z} \left [  \frac{\gamma H}{1-\gamma(1+H^{\prime})} \frac{\partial p'}{\partial z} \right ] - 
$$
$$-\left[ \frac{g k^2}{\sigma^2} + \frac{\gamma(1+H^{\prime})^2}{4H[1-\gamma(1+H^{\prime})]} - \frac{\gamma H^{\prime \prime}}{2[1-\gamma(1+H^{\prime})]^2} \right] p' +$$
\begin{equation}
  +  \frac{i g}{\sigma}\left[\frac{\partial}{\partial y} - \frac{k}{\sigma} f\right] v_y = 0,
  \label{Eq.(14)}
 	\end{equation}
where $H^{\prime}$ means the derivative of $H$ by $z$.

We eliminate $p'$ from Eqs. (\ref{Eq.(13)})-(\ref{Eq.(14)}) and derive the single equation for $v_y$\\
$$ \frac{\partial }{\partial z} \left [  \frac{\gamma H}{1-\gamma(1+H^{\prime})} \frac{\partial v_y}{\partial z} \right ] - 
$$
$$-\left[\frac{\gamma(1+H^{\prime})^2}{4H[1-\gamma(1+H^{\prime})]} - \frac{\gamma H^{\prime \prime}}{2[1-\gamma(1+H^{\prime})]^2} \right]v_y-$$

\begin{equation}
 - \frac{g}{f^2 - \sigma^2} \left(\frac{\partial^2}{\partial y^2} - k^2 - \frac{k}{\sigma} \frac{d f}{dy}\right) v_y = 0.
 \label{Eq.(15)}
 	\end{equation}
This equation can be solved by the separation of variables in appropriate boundary conditions. Here we note that the separation of variables is only possible in the case of vertically hydrostatic assumption.  

We represent $v_y$ as 
\begin{equation}
	v_y=V(z) \Psi(y)
	\label{Eq.(16)}
	\end{equation}
and after straightforward calculations with the separation constant $-{h^{-1}}$ we derive the two equations
\begin{equation}
	\frac{\partial^2 \Psi}{\partial y^2} + \left[\frac{\sigma^2 -f^2}{g h} - k^2 - \frac{k}{\sigma}\frac{d f}{dy} \right ] \Psi = 0,
	\label{Eq.(17)}
	\end{equation}
	
	$$ \frac{\partial }{\partial z} \left [  \frac{\gamma H}{1-\gamma(1+H^{\prime})} \frac{\partial }{\partial z} \right ]V(z) - 
$$
\begin{equation}
-\left[ \frac{\gamma(1+H^{\prime})^2}{4H[1-\gamma(1+H^{\prime})]} - \frac{\gamma H^{\prime \prime}}{2[1-\gamma(1+H^{\prime})]^2} + \frac{1}{h} \right]V(z) = 0.
\label{Eq.(18)}
	\end{equation}

 Here Eq. ~(\ref{Eq.(17)}) and Eq. ~(\ref{Eq.(18)}) are the latitudinal and vertical equations governing the latitudinal and vertical structures of the waves, correspondingly. Actually, Eq. ~(\ref{Eq.(17)}) is equivalent to the equation, which governs shallow water equatorially trapped waves in a homogeneous layer with the width of $h$ (e.g. \citet{Matsuno1966}). This conclusion corresponds to the Taylor theorem \citep{Taylor1936}, which states that the dynamics of Rossby waves in stratified fluids is identical to the waves in a homogeneous layer, which has a width of equivalent depth. The equivalent depth is different for different modes of Rossby waves. The theorem is valid for all compressible and stratified fluids \citep{Pedlosky1987,Zaqarashvili2021}. We first solve Eq. ~(\ref{Eq.(18)}) appropriate boundary conditions and find the equivalent depth, $h$. Then we use it to solve Eq. ~(\ref{Eq.(17)}) and find the solutions in $y$ direction satisfying bounded boundary conditions at poles. It should be noted that the value of equivalent depth, $h$, found from Eq. ~(\ref{Eq.(18)}), defines the solutions of Eq. ~(\ref{Eq.(17)}), therefore the two equations are not independent but interconnected by $h$.

Before starting to study the case of inhomogeneous distribution of temperature with depth, we present the solutions of the simplest case of uniform $H$, which means an isothermal temperature profile. This is a very simplified approach, but it gives the basic physics of oscillations.

With $H=const$ Eq.~(\ref{Eq.(18)}) leads to
\begin{equation}
    \frac{\partial^2 V(z)}{\partial z^2}-\left[\frac{1}{4H^2}- \frac{\kappa}{ H h}\right] V(z) = 0,
    \label{Eq.(19)}
\end{equation}
where $\kappa=(\gamma - 1)/\gamma$.
	
Solution of Eq.~(\ref{Eq.(19)}) is either exponential or periodic functions depending on the value of the equivalent depth, $h$. We consider the exponential function for close boundary condition, $v_z$=0 at the surface, which corresponds to $z=0$. The value of scale height depends on the surface temperature. The surface temperature of 10 000 K gives $H=300$ km and the corresponding equivalent depth found from the close boundary condition is $h \approx$ 500 km.

The equivalent depth can be used to solve the latitudinal equation, Eq. ~(\ref{Eq.(17)}). The solutions of Eq. ~(\ref{Eq.(17)}) must satisfy boundary conditions in $y$ directions, namely they must exponentially vanish at poles. Only the equivalent depth which satisfies the polar boundary conditions can be considered to be valid. The solution of the latitudinal equation for $h=500$ km does not satisfy boundary conditions at poles, therefore the exponential solution of Eq.~(\ref{Eq.(19)}) is not valid. The periodic solution of Eq.~(\ref{Eq.(19)}) leads to the value of equivalent depth, which satisfies the polar boundary conditions, but it corresponds to very high values of vertical wavenumber.
   
  When the temperature is a function  of depth (generally increasing), then its  gradient governs the state of the medium, which could be adiabatic, radiative or convective. The Ledoux function, $A$, which defines the state, can be written using Eqs. ~(\ref{Eq.(4)})-(\ref{Eq.(6)}) in terms of the density scale height as 
  \begin{equation}
  A= \frac{R}{\rho_0}\frac{d\rho_0}{d z} - \frac{R}{\gamma p_0}\frac{d p_0}{d z}=-\frac{R}{H}\left (\frac{\gamma-1}{\gamma}+\frac{d H}{dz} \right ).
  \label{Ledoux}
     \end{equation}
  
 When $|dH/dz|=|H^{\prime}|>{(\gamma-1)}/{\gamma}$, i.e. $A>0$,  the temperature gradient is super-adiabatic and corresponds to convective stars. When
 $|H^{\prime}|<{(\gamma-1)}/{\gamma}$, i.e. $A<0$, the temperature gradient is sub-adiabatic, hence it corresponds to radiative stars. When $A=0$ i.e. $|H^{\prime}|=\kappa={(\gamma-1)}/{\gamma}=2/5$, the star is neutrally stable so that the temperature gradient is adiabatic, which is a limiting case of super- and sub-adiabatic gradients.
 In this paper, we consider radiative stars, therefore the condition of $|H^{\prime}|<\kappa$ should be satisfied everywhere. This condition is most easily satisfied for the linear profile of temperature with an uniform vertical gradient. For other profiles, the vertical temperature gradient generally increases with depth, which unavoidably leads to the violation of the radiative condition at some distance from the surface. Therefore, we assume that the density scale height is a linear function of depth (remember that $z>0$ above the surface and $z<0$ below the surface) i. e.  
   \begin{equation}
 H=H_0-\epsilon z.
  \label{H}
     \end{equation}
   This is equivalent to the temperature profile of the form
        \begin{equation}
  T=T_0 \left (1- \epsilon \frac{z}{H_0} \right ),
  \label{Eq.(20)}
     \end{equation}
  where $T_0$ is the temperature at the surface, $z=0$.  In this case, the radiative stars imply $\epsilon<\kappa$ and we will use the criterion throughout the paper.

    \begin{figure*}
   \centering
   \includegraphics[width=20cm]{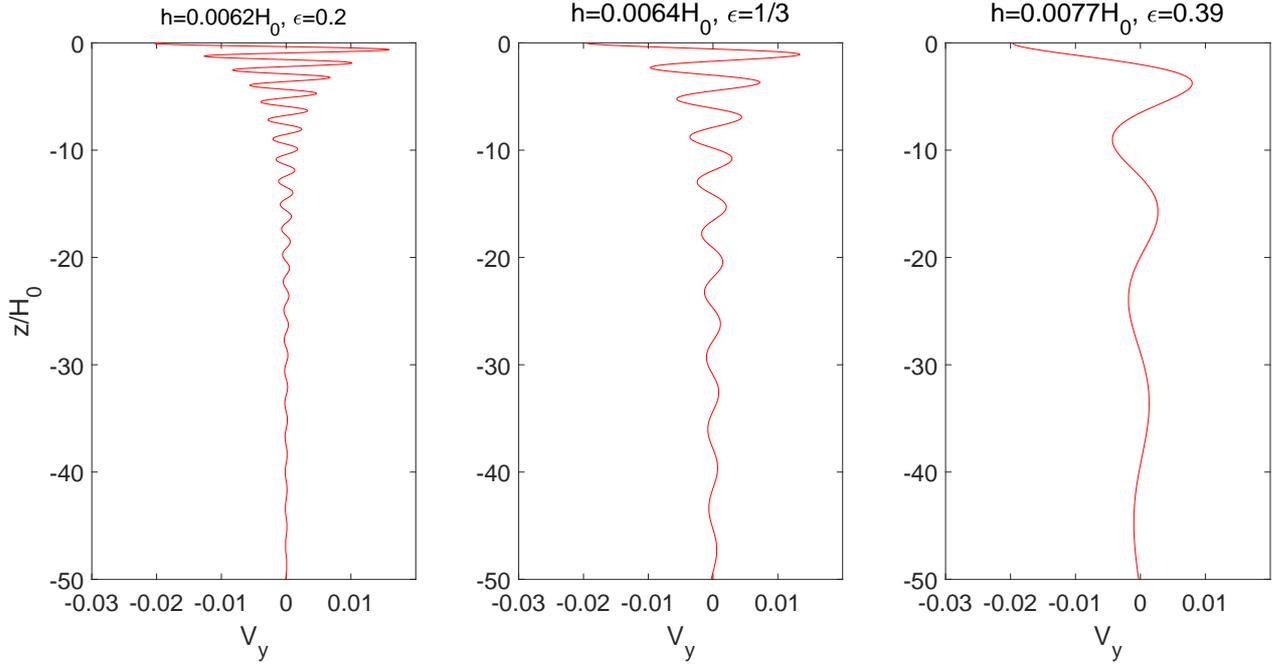}
\caption{Vertical structure of Rossby waves for different vertical temperature gradients: $\epsilon=0.2$  (left panel), $\epsilon=1/3$  (middle panel), $\epsilon=0.39$ (right panel). Here $v_y$ is plotted without $\sqrt{\rho_0}$ in Eq.~(\ref{Eq.(7)}) and normalised by $\Omega R$. 
              }
         \label{Fig.1}
  \end{figure*}

  %Free and forced oscillation cases will be considered in separate sections.
  
   \section{Free oscillations of radiative stars}
 
 We start to study free oscillations, which means to solve first the vertical structure equation ~(\ref{Eq.(18)}) in corresponding boundary conditions.  The solutions of the equation govern the spatial structure of Rossby waves along the vertical direction. The boundary conditions allow us to find the discrete values of equivalent depth, $h$. Then we will use $h$ to find latitudinal solutions of Eq. ~(\ref{Eq.(17)}), which exponentially tend to zero at poles.  These solutions correspond to the latitudinal structure of Rossby waves.

\subsection{Vertical structure of Rossby waves}

 For the linear temperature profile, Eq. ~(\ref{Eq.(18)}) is rewritten as
       \begin{equation}
      \frac{\partial}{\partial z} \left[ \frac{\gamma H}{1-\gamma (1+H')} \frac{\partial}{\partial z} \right] V(z)-\left[ \frac{\gamma (1+H')^2}{4H [1-\gamma (1+H')]} + \frac{1}{h} \right] V(z) = 0,
      \label{Eq.(21)}
   \end{equation}
   where $H'=d H/dz$. Using the new variable
   \begin{equation}
       x=2\sqrt{H_0-\epsilon z} \frac{\sqrt{\gamma (1-\epsilon )-1}}{\epsilon \sqrt{\gamma h}}
       \label{Eq.(22)}
   \end{equation}
Eq. ~(\ref{Eq.(21)}) leads to
\begin{equation}
x^2\frac{\partial ^2 V(x)}{\partial x^2} + x \frac{\partial V(x)}{\partial x}+ ( x^2 - n^2) V(x) = 0,
\label{Eq.(23)}
\end{equation}
\\
where $n=(1-\epsilon)/\epsilon$. This is the Bessel equation  and its solutions are Bessel functions of order $n$, $J_n(x)$, and $Y_n(x)$. The solutions $J_n(x)$ and $Y_n(x)$ must satisfy certain boundary conditions at the surface. 
We use two different boundary conditions.  The first condition yields that the vertical velocity vanishes at the surface, i.e. $v_z=0$ at $z=0$, which is a close condition. The second condition yields that the total Lagrangian pressure is zero at the surface, which is a free boundary condition. In both cases, vertical velocity and total pressure must be bounded towards the stellar center.

We use different values of $\epsilon$ to find the vertical structure of Rossby waves for different vertical temperature gradients. We assume $\epsilon=0.2, 1/3, 0.39$, which give the order of Bessel functions as $n=4, 2, 1.56$, correspondingly. Note that the adiabatic temperature gradient corresponds to $\epsilon=0.4$, therefore $\epsilon=0.39$ is nearly upper limit of radiative temperature gradient.

\subsubsection{Close boundary condition, $v_z$=0, at the surface}

The close boundary condition, $v_z$=0, from Eqs.~(\ref{Eq.(10)})-(\ref{Eq.(13)}) yields the following equation 
   \begin{equation}
       \frac{\partial v_y}{\partial z} + \left[ \frac{1}{\gamma H} - \frac{1}{2H} - \frac{H'}{2H} \right] v_y=0.
       \label{Eq.(24)}
   \end{equation}
  Both solutions of Eq.~(\ref{Eq.(23)}), $J_n(x)$ and $Y_n(x)$, have the similar vertical structures of the modes. Therefore, we take $J_n(x)$ as a solution and then Eq.~(\ref{Eq.(24)}) is rewritten as \\
   \begin{equation}
       \frac{\partial}{\partial x} J_n(x) + \frac{1}{x} \left[ \frac{1}{\epsilon} - 1 - \frac{2}{\gamma \epsilon } \right] J_n(x) = 0.
       \label{Eq.(25)}
   \end{equation}
   \\
   This is a transcendental equation, which has the infinite number of zeros. Each zero defines a certain value of equivalent depth, $h$, hence corresponds to a certain wave mode. We first solve the equation for $\epsilon=1/3$ and
   for the first 6 zeros, we obtain the values of equivalent depth as $h \approx 1.05 \, H_0 $, $h \approx 0.058 \, H_0 $, $h \approx 0.025 \, H_0$, $h \approx 0.014 \, H_0 $, $h \approx 0.009 \, H_0 $, and $h \approx 0.0064 \, H_0 $ respectively. As we mentioned above, only the values of equivalent depth (or wave modes), which result in the bounded conditions at poles, are valid. Solutions of the latitudinal equation (see the next subsection) show that the modes corresponding to the first five zeroes do not satisfy polar boundary conditions and hence are not valid. On the other hand, the modes corresponding to the sixth (and larger) zero satisfy bounded conditions at poles. The solutions of  Eq.~(\ref{Eq.(25)}) for the temperature gradient of  $\epsilon=0.39$ show that the mode, which corresponds to the second zero of the equation, already satisfies the polar boundary conditions. Therefore, all modes starting from the second mode in the vertical direction are valid for a nearly adiabatic temperature gradient. On the other hand, the temperature gradient of  $\epsilon=0.2$ yields that 17th and higher modes give the bounded polar boundary conditions. Figure~ \ref{Fig.1} shows the vertical structure of Rossby waves based on the solutions of Eq.~(\ref{Eq.(23)}) for different values of $\epsilon$.
   We see that all solutions exponentially decrease with depth, hence the corresponding modes are trapped near the surface. All higher modes show similar behavior. The figure shows that the smaller values of $\epsilon$, i.e. the smaller temperature gradient, yield the shorter vertical wavelength of Rossby waves and the stronger decay of the wave amplitude with depth. Therefore, the Rossby waves tend to concentrate nearer to the surface for a smaller temperature gradient.
   \\

\begin{figure}[t]
    \centering
    \includegraphics[width=0.5\textwidth]{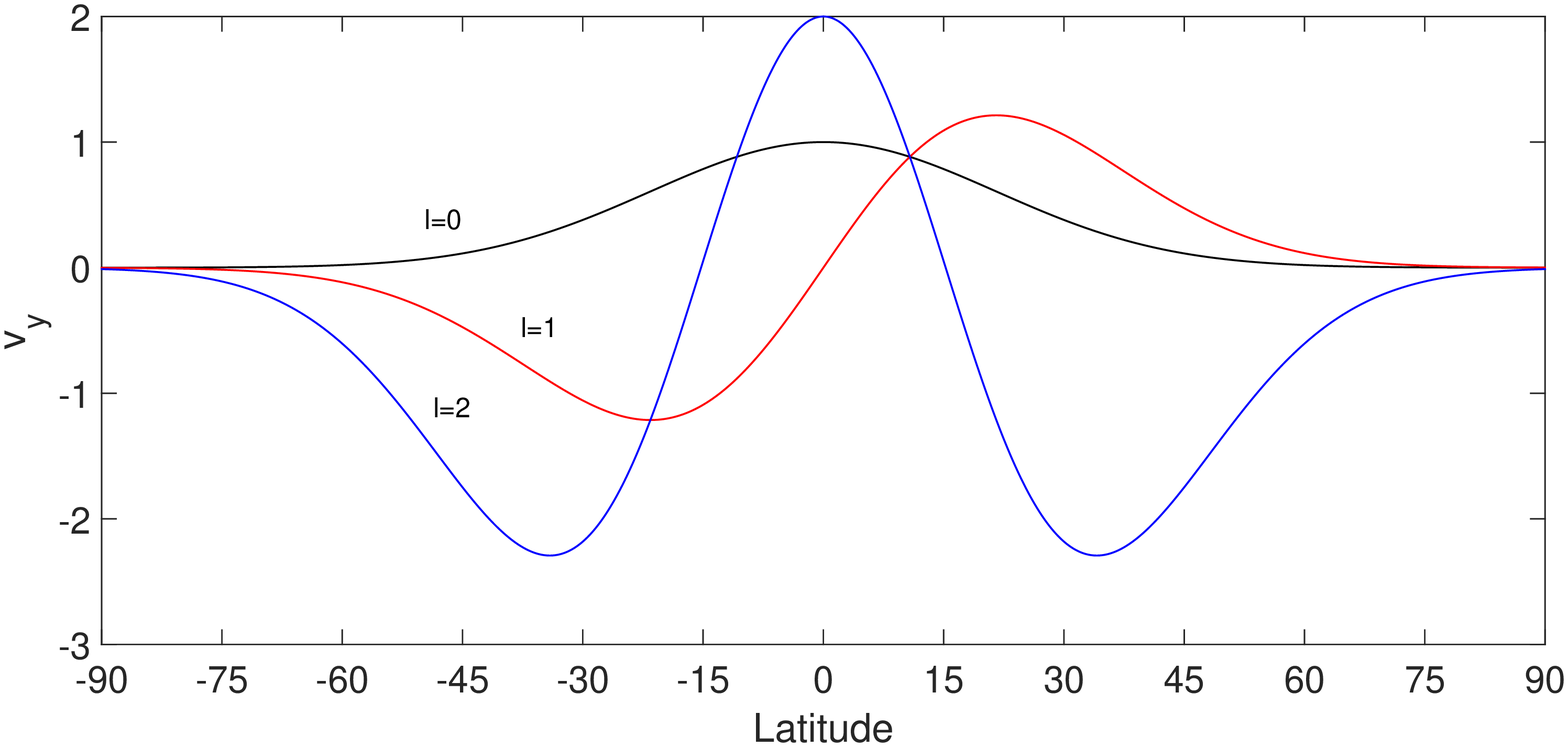}
    \caption{Latitudinal structure of equatorially trapped Rossby waves for the temperature gradient of $\epsilon=1/3$ {and the value of equivalent depth $\approx 0.0064 H_0$}. The solutions exponentially decay towards poles above middle latitudes. Black, red and blue curves show $l=0$, $l=1$ and $l=2$ modes. }
    \label{Fig.3}
\end{figure}

 %-----------------------------------------
   \subsubsection{Free boundary condition}
   In the free surface condition, the total Lagrangian pressure is zero at the surface. Using Eqs. ~(\ref{Eq.(10)}) and (\ref{Eq.(12)}), the condition is  
 
   \begin{equation}
      \frac{\partial v_y}{\partial z} + \frac{1}{H}  \frac{3(1- \epsilon)}{2}  v_y = 0,
      \label{Eq.(27)}
   \end{equation}
  which can be rewritten for the new variable $x$ as\\
   \begin{equation}
      \frac{\partial J_n (x)}{\partial x} - \frac{1}{x}\frac{3(1-\epsilon)}{\epsilon} J_n(x) = 0
      \label{Eq.(28)}
   \end{equation}
   This is also a transcendental equation and has the infinite number of zeros. We assume $\epsilon=1/3$ and for the first five zeroes we found $h \approx 0.069 \, H_0$,  $h \approx 0.027 \, H_0$,  and $h \approx 0.015\, H_0$, $h \approx 0.0094 \, H_0$, $h \approx 0.0065 \, H_0$, respectively. Only the modes corresponding to the fifth (and higher) zero satisfy the polar boundary conditions.  

  { The value of $h$ obtained from each zero of free condition is similar to the value obtained from the next higher zero of close condition. For example, the first zero of free condition is near to the second zero of close condition, etc. This happens because of the relation between total Lagrangian pressure and vertical velocity.  Therefore, close and free boundary conditions result in similar spatial structure with depth of the corresponding modes. Consequently, we will only consider the closed boundary condition in the rest of the paper. }
   
   \subsection{Latitudinal structure of Rossby waves}
   
   We now turn to the latitudinal equation (Eq.~(\ref{Eq.(17)})) and find the solutions with certain $h$ satisfying bounded conditions at poles, which define dispersion relations of possible wave modes in the system.

     { The solutions of Eq.~(\ref{Eq.(17)}) crucially depend on the parameter
     \begin{equation}
 \varepsilon=\frac{4\Omega^2 R^2}{g h},	\label{Lamb}
	\end{equation}
which actually corresponds to the parameter (sometimes called as Lamb parameter) governing the dynamics of shallow water system of the layer thickness, $h$. 
When this parameter is much larger then 1 (for fast rotation or small $h$), then Eq.~(\ref{Eq.(17)}) is most easily satisfied for small $y$. \footnote{see e. g. \citet{Longuet-Higgins1968} for the similar consideration in the spherical geometry.}  Small $y$ is equivalent to the equatorial region and hence the solutions are equatorially trapped. Third and higher zeroes of close boundary condition with $h < 0.025 \, H_0$ lead to the Lamb parameter of  $\varepsilon > 10$ for solar radius, surface temperature and rotation. Therefore, the modes corresponding to the higher zeroes of close boundary condition are confined near the equatorial regions and decay sufficiently fast towards the poles. The modes, hence, can be considered by the equatorial beta-plane approximation, which means to expand the Coriolis parameter near the equator $\theta \approx 0$ and retain only the first order term, $f=\beta y$, where $\beta = (2 \Omega/R)$. }
In this case Eq.~(\ref{Eq.(17)}) tends to
\begin{equation}
	\frac{\partial^2 \Psi}{\partial y^2} + \left[-\frac{k \beta}{\sigma} - k^2 + \frac{\sigma^2}{c^2} - \frac{\beta^2 y^2}{c^2} \right] \Psi = 0,
	\label{Eq.(29)}
	\end{equation}
where $c=\sqrt{g h}$ is the surface gravity speed for corresponding equivalent depth, $h$, which is obtained from the vertical structure equation as shown in the previous subsections. \\
   
    { It must be noted that} the solutions under equatorial beta-plane approximation fairly correspond to the solutions of spherical case. In fact, the governing equation of equatorially trapped Rossby waves in  beta-plane approximation is identical to the spherical case \citep{Longuet-Higgins1968, Zaqarashvili2021}. Here we consider only the equatorially trapped waves, for which the beta-plane approximation is justified.
    
Eq. (\ref{Eq.(29)}) is a parabolic cylinder equation which has bounded solutions when
\begin{equation}
    -\frac{k \beta }{\sigma} - k^2 + \frac{\sigma^2}{c^2} = \frac{\beta}{c} (2l+1),
    \label{Eq.(30)}
\end{equation}
where $l=0, 1, 2, 3 ...$. Then the solution to this equation is
\begin{equation}
    \Psi=\Psi_0 exp\left[-\frac{\beta}{2c}y^2 \right] H_l \left (\sqrt{\frac{\beta}{c}}y \right )
    \label{Eq.(31)}
\end{equation}
where $\Psi_0$ is the value of $v_y$ at the equator, $y=0$, and $H_l$ is the Hermite polynomial of the order $l$. The solution is oscillatory inside the interval $y< |\sqrt{(2l+1)c/\beta}|=|\sqrt{(2l+1)R^2/\sqrt{\varepsilon}}|$ and exponentially decreases towards poles outside. Smaller $h$ or larger $\varepsilon$ yields the stronger decrease of the solution. { As it was discussed above,} the values of $h$ corresponding to the lower zeroes of Eqs. (\ref{Eq.(25)}) and (\ref{Eq.(28)}) lead to solutions which are not bounded at poles. The bounded solutions in the case of $\epsilon=1/3$, which are shown on Figure ~\ref{Fig.3}, start to appear for the sixth (fifth) zero of close (free) boundary condition, which yields $h \approx 0.0064 \, H_0$. In this case, the Lamb parameter is around $\varepsilon \approx 43$ for solar radius, surface temperature and rotation. { The temperature gradient of $\epsilon=0.39$ affords that the second { (and higher)} zero of close vertical boundary condition leads to the bounded solution along latitudes. The critical latitude is estimated as $\theta_c=\arcsin \left ( \sqrt{(2l+1)/\sqrt{\varepsilon}}\right )$, which for $l=1$ modes gives around $\pm 40^0$, i.e. the solutions are oscillatory in the latitudes of $<40^0$ and exponentially decay for $>40^0$. Therefore, the solutions are  mostly concentrated between the latitudes $\pm 60^0$ and are negligible at poles satisfying the boundary conditions there.

\begin{figure}[t]
    \centering
    \includegraphics[width=0.5\textwidth]{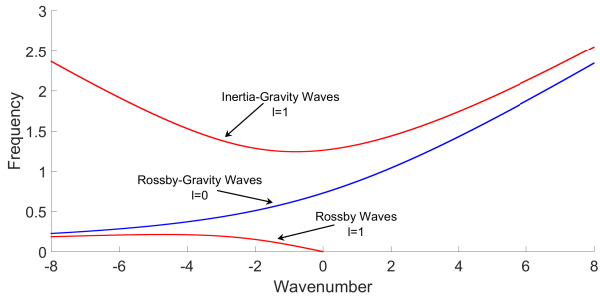}
    \caption{The solutions of full dispersion equation (Eq.~(\ref{Eq.(32)})) for $\varepsilon=43$, which corresponds { to the first valid vertical mode (with $h=0.0064 H_0$) in a star with the rotation, radius and surface gravity of the Sun and the temperature gradient of $\epsilon=1/3$.} Red curves correspond to the inertia-gravity (upper) and Rossby (lower) waves with $l=1$. The blue curve corresponds to the Rossby-gravity waves with $l=0$. Wave frequency is normalized by the angular frequency of the star, $\Omega$. Toroidal wavenumber, $k$, is normalized by the stellar radius, $R$.}
    \label{Fig.4}
\end{figure}

\subsection{Dispersion equation}

Eq.~(\ref{Eq.(30)}) defines the dispersion relation of waves
\begin{equation}
\sigma^3- 4\Omega^2 \left (\frac{k^2R^2}{\varepsilon}+\frac{2l+1}{\sqrt{\varepsilon}}\right ) \sigma -8\Omega^3 \frac{k R}{\varepsilon} =0;
\label{Eq.(32)}
\end{equation}
This equation is identical to the dispersion relation of shallow water waves when the width of shallow layer is replaced by the equivalent depth, $h$, as stated by the Taylor theorem. The dispersion relation governs the inertia-gravity, Rossby and Rossby-gravity waves for each $h$. 

For the high frequency limit ($\sigma \gg \Omega$) with $l \geq 1$, we have the dispersion relation
\begin{equation}
\sigma=\pm 2\Omega \sqrt{\frac{k^2R^2}{\varepsilon}+\frac{2l+1}{\sqrt{\varepsilon}}},
\label{Eq.(33)}
\end{equation}
which corresponds to the inertia-gravity waves.

For the low frequency limit ($\sigma \ll \Omega $) with $l \geq 1$, we have the dispersion relation
\begin{equation}
 \sigma= -2\Omega\frac{kR}{k^2 R^2+(2l+1)\sqrt{\varepsilon}},
 \label{Eq.(34)}
\end{equation} 
which corresponds to the Rossby waves.

For $l=0$, Eq.~(\ref{Eq.(32)}) leads to  
\begin{equation}
\left (\sigma-2\Omega\frac{k R}{\sqrt{\varepsilon}} \right )\left (\sigma^2-2\Omega\frac{k R}{\sqrt{\varepsilon}}\sigma -\frac{4\Omega^2}{\sqrt{\varepsilon}} \right ) =0.
\label{Eq.(35)}
\end{equation}
First solution of Eq.~(\ref{Eq.(34)}) is spurious, therefore must be neglected and the second solution defines the Rossby-gravity wave
\begin{equation}
\sigma=\Omega\frac{k R}{\sqrt{\varepsilon}}\left (1 \pm \sqrt{1+\frac{4\sqrt{\varepsilon}}{k^2R^2}} \right ).
\label{l=0}
\end{equation}

The solutions of full dispersion equation (Eq.~(\ref{Eq.(32)})) are displayed on Figure~\ref{Fig.4} { for ${\varepsilon}=43$, which corresponds to the equivalent depth, $h \approx 0.0064 \, H_0 $ (associated with the first valid vertical mode or the sixth zero of close condition for the temperature gradient of $\epsilon=1/3$) and a star with the rotation, radius and surface gravity of the Sun.} 
%Red curves correspond to the inertia-gravity (upper curve) and the Rossby (lower curve) modes with $l=1$. Blue curve corresponds to the mixed Rossby-gravity mode with $l=0$. 
We see that for large negative $k$, the dispersion curves of Rossby and Rossby-gravity waves merge, while for the large positive $k$ the inertia-gravity and Rossby-gravity waves have the same behavior. In the next subsection we will consider Rossby and Rossby-gravity waves in detailed.

\subsection{Rossby and Rossby-gravity waves}

Eq.~(\ref{Eq.(34)}) is the dispersion relation for the Rossby waves on the equatorial beta-plane. The positive frequency and the negative toroidal wavenumber indicate to the retrograde (opposite to the rotation) propagation of waves. The dispersion relation crucially depends on the parameter $\varepsilon$. When this parameter is small (i.e. for slowly rotating stars or large equivalent depth, $h$) then the second term in denominator is small than the first one, which eventually leads to the dispersion relation of Rossby waves on the 2D surface with only longitudinal propagation ($k_y=0$). This corresponds to sectoral harmonics in the spherical geometry. But moderate value of the parameter significantly changes the dispersion relation of Rossby waves. In the considered case, this parameter has a moderate value due to the small $h$. For a star with solar parameters, i.e. $T_0=5770$ K, $\Omega=3 \times 10^{-6}$ s$^{-1}$, $R=7\times 10^5$ km, and $h=0.0064 \, H_0$ km as estimated in previous subsection, one can find that $\varepsilon \approx 43$. Therefore, one must keep the associated term in the dispersion relation. One should note that the parameter $\varepsilon$ is equivalent to the parameter used by \citet{Provost1981} to study the r-modes in slowly rotating stars, namely $\Omega^2 R^3/(G M)$, if one replaces $h$ by $R$ and divide by 4.  \citet{Provost1981} expanded all variables and the frequency with this parameter assuming it much smaller than unity. This could be correct if the equivalent depth, $h$, is of the order of stellar radius. But our solution shows that $h\ll R$, i.e. the expansion parameter of \cite{Provost1981} is not small and can not be appropriate in our consideration. 

Rossby-gravity waves ($l=0$) defined by Eq.~(\ref{l=0}) have mixed properties of Rossby and inertia-gravity waves. This mode is similar to Rossby waves when it propagates opposite to the rotation, but it reminds the gravity wave when propagates to the direction of the rotation. The wave has no oscillatory solution along latitudes and hence it corresponds to the sectoral modes in the spherical geometry. 

\section{Observational constraints}

It is of importance to show how the theoretically obtained waves might be seen by observations. Recent progress in observations of Rossby waves using Kepler light curves suggest that the waves can be also detected in the data of TESS mission. Here we provide hints for observers to detect the waves. 

  \begin{figure*}
   \centering
   \includegraphics[width=18cm]{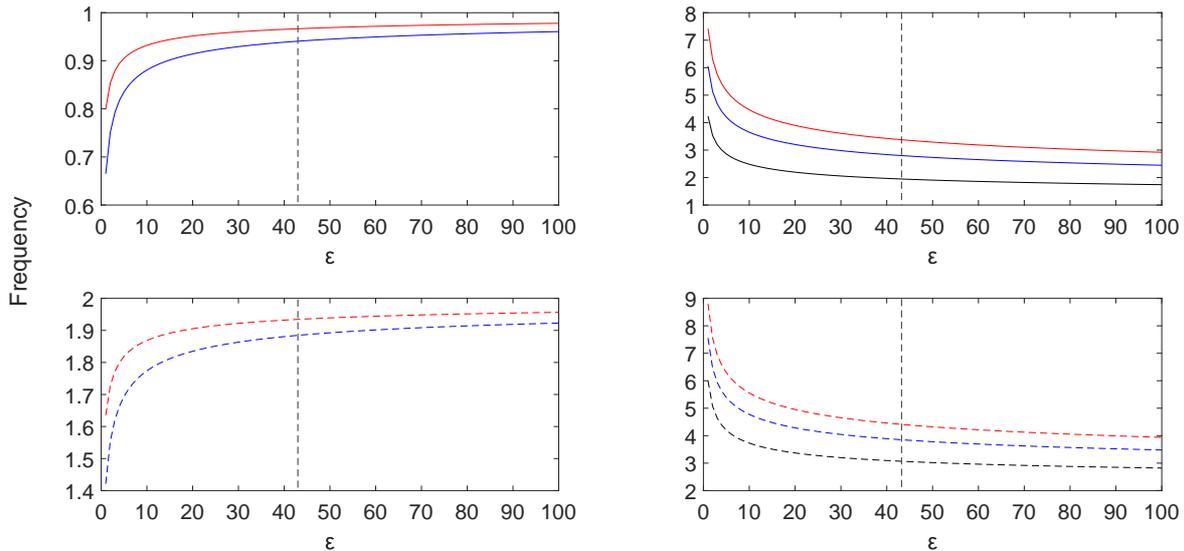}
\caption{Wave frequency in the inertial frame as expected to be seen by observations vs $\varepsilon$. The frequency is normalized by the angular frequency of a star, $\Omega$. Black, blue and red curves correspond to the modes with $l=0$, $l=2$, and $l=4$, respectively. The left panels display the Rossby modes with $m=1$ (solid lines, upper panel) and $m=2$ (dashed lines, lower panel). Upper (lower) right panel shows inertia-Gravity modes (Rossby-Gravity modes). Grey dashed lines denote the value of $\varepsilon$ calculated for the solar parameters and the equivalent depth of $h=0.0064\, H_0$. 
             }
        \label{Fig.5}
   \end{figure*}

We used the rotating frame for the theoretical analysis of Rossby waves. On the other hand, observed light curves are obtained in the inertial frame, therefore the observable frequency of the waves is expressed by
\begin{equation}
\sigma_{obs}=m \Omega + \sigma,
 \label{obs}
\end{equation}
where $\sigma$ is the theoretical wave frequency in the rotating frame and $m=k R$ is the normalized toroidal wavenumber.

Wave frequency depends on the parameter, $\varepsilon$, which can be rewritten as
\begin{equation}
 \varepsilon=0.3 \left (\frac{h}{H_0} \right)^{-1} \left (\frac{T_0}{T_{sun}} \right)^{-1} \left( \frac{\Omega}{\Omega_{sun}} \right)^{2} \left( \frac{R}{R_{sun}} \right)^{2},	\label{epsilon}
\end{equation}
where $T_{0}$ ($T_{sun}$), $\Omega$ ($\Omega_{sun}$) and $R$ ($R_{sun}$) are the surface temperature, the surface angular velocity and the radius of a star (the Sun), respectively.

Figure \ref{Fig.5} shows the dependence of wave frequency in the inertial frame, $\sigma_{obs}$ on $\varepsilon$. Only the frequency of lower order modes $m=1$ and $m=2$ are shown on this figure. The normalized frequencies of all modes (Rossby, Rossby-gravity, inertia-gravity) are increasing for higher $\varepsilon$ i.e. for higher angular velocity of stars  (with same surface temperature and radius). The anti-symmetric harmonics with regards to the equator, $l=1, 3...$, probably have negligible contributions to stellar light curves as the northern and southern parts of the modes will cancel each other.\footnote{This applies to stars those rotation axes are perpendicular to the line of sight. For the stars with inclined rotation axes, the things are more complicated.} Therefore, only symmetric harmonics, $l=0, 2, 4$, are shown on this figure.  Rossby waves with $m=1$ have the frequency in the range of $0.65 \, \Omega < \sigma_{obs}<\Omega$ i.e. less than the stellar angular velocity. The frequency of Rossby waves with $m=2$ is in the range of $1.4 \, \Omega < \sigma_{obs}<2 \Omega$. Inertia-gravity and Rossby-gravity waves have an order of magnitude higher frequencies than the angular velocity of stellar rotation. Observations can define the frequency of observed waves, which will determine the corresponding value of $\varepsilon$ using Figure \ref{Fig.5}. { $\varepsilon$ depends on the equivalent depth, $h$, and three stellar parameters such as the surface temperature, the surface angular velocity and the radius. The equivalent depth is calculated from the Rossby wave theory as discussed in the subsection 3.1. Consequently, if one knows two of the stellar parameters can estimate the value of the third one. }

\section{Discussion and conclusions}

Rossby waves arise due to the conservation of absolute vorticity, therefore the vorticity is an essential ingredient of the waves. On the other hand, the pressure variation is an equally important component in the waves. Therefore, the waves will cause the periodic variations of surface pressure (temperature or density or both), which may lead to the periodical modulation of stellar radiance \citep{Saio2018}. Recent observations of Rossby waves in the light curves of Kepler and TESS stars \citep{Saio2018,Li2019,Jeffery2020,Samadi2020, Takata2020,Henneco2021,Saio2022} opens a new area for seismology of stellar interiors by the waves.

It is important to know the oscillation spectrum of the modes and how deeper the modes penetrate in the stellar interior. To solve the full spherical 3D problem is generally complicated also with numerical simulations. In this paper, we look into more simplified rectangular problem taking into account the vertical stratification of density and temperature in stellar interiors. The rectangular geometry significantly simplifies the finding of wave dispersion relations, which are similar to those obtained under the spherical geometry \citep{Matsuno1966,Longuet-Higgins1968}. Previous theoretical studies generally considered the spherical geometry in slowly rotating limit \citep{Provost1981,Saio1982,Damiani2020}. Though, the approximation of \citet{Papaloizou1978} includes rapidly rotating stars, it is valid for only higher order modes. On the other hand, our approximation is valid for the stars with any rotation rate (except very rapidly rotating stars with significant distortion from the spherical symmetry), therefore it is step forward in the study of stellar Rossby waves.  Our mathematical formalism closely follow to the consideration adapted in the Earth atmosphere \citep{Lindzen1967}. We considered the vertically hydrostatic assumption, so that the vertical distribution of the pressure is only slightly disturbed from its static form due to the waves. This approximation neglects the internal gravity and acoustic waves, hence it is valid for time and spatial scales of Rossby waves. For linear dynamics of the waves,  we derived single second order partial differential equation for vertical and latitudinal variations, which is solved by the method of separation of variables. The two equations for the vertical and the latitudinal variations are obtained, which are connected by the separation constant defining the equivalent depth, $h$. The solutions of the two equations which satisfy certain boundary conditions give the exact analytical dispersion relations and the vertical structure of the waves in certain distribution of background values.

Vertical structure of wave modes obviously depends on the vertical temperature profile. The super-adiabatic temperature gradient (positive Ledoux function, $A>0$) corresponds to convective stars, while the sub-adiabatic gradient (negative Ledoux function, $A<0$) describes the radiative stars. Limiting case from both gradients i.e. $A=0$ is the adiabatic temperature gradient, which fits the neutrally stable interior, where any plasma displacement in the vertical direction has no following dynamics. In this paper, we consider only radiative stars with the linear sub-adiabatic temperature gradient. \footnote{Convective stars with near super-adiabatic temperature gradient will be studied in a forthcoming paper.} Other profiles of temperature lead to the changing of Ledoux function from the negative to the positive value at some depth, therefore they are inappropriate for the radiative stars. In the case of the linear temperature profile, the vertical structure equation is transformed into the Bessel equation, which has exact analytical solutions in terms of Bessel functions.

To solve the latitudinal equation, we use the equatorial beta-plane approximation, which resulted in a parabolic cylinder equation with known solutions in terms of Hermite polynomials.  Any solution on the equatorial beta-plane, which decays sufficiently fast towards the coordinate corresponding to the pole, is a correct approximation to the solutions on a sphere \citep{Lindzen1967}. Indeed, the governing equation and the dispersion relation of equatorially trapped waves are identical in the equatorial beta-plane and spherical geometry \citep{Longuet-Higgins1968, Zaqarashvili2021}. Hence, the beta-plane approximation is valid for the solutions decaying towards the poles. The solutions satisfying bounded boundary conditions at poles defined dispersion equation for Rossby, Rossby-gravity and inertia-gravity waves, Eq. (\ref{Eq.(32)}). Frequencies of modes with different wave numbers are then easily derived.    

The solutions of the problem significantly depend on boundary conditions for vertical and latitudinal structure equations. We first solved the vertical structure equation in close boundary condition at the surface (i.e. when the vertical velocity vanishes), which led to the equivalent depth, $h$. Then we used the depth to find the solutions of latitudinal equation, which satisfied the bounded conditions at poles. 

Close boundary conditions at the surface led to the transcendental equation with Bessel functions (Eq. \ref{Eq.(25)}), which has infinite number of zeroes  corresponding to different wave modes. Each of the zeroes define the equivalent depth, which shapes the latitudinal structure of modes, therefore only the modes which have bounded solutions at poles are valid.  We found that the first valid zero yields the equivalent depth of $h=0.0064 \, H_0$ for the vertical temperature gradient of $\epsilon=1/3$, where $H_0$ is the density scale height at the surface. In this case, the modes with $l=0, 1, 2$, where $l$ shows the number of zeroes between the poles, are concentrated around the equator between $\pm 60^0$ latitudes (see Fig. 2), so they are the equatorially trapped waves \citep{Matsuno1966,Longuet-Higgins1968}. The modes have oscillatory behavior along the vertical direction with the wavelength of several surface scale heights (see  middle panel on Figure 1) and may penetrate to the depth of $\sim 50 \, H_0$ (the scale height of a star with the surface temperature of 10 000 K and solar-type surface gravity is around 300 km).
 The vertical structure of modes significantly depends on the vertical temperature gradient rate. We found that the vertical wavelength of modes is longer for the stronger temperature gradient (Figure 1) being of the order of density scale height for $\epsilon=0.2$ and of the order of 10 density scale height for $\epsilon=0.39$. Figure 1 also shows that the smaller temperature gradient leads to the stronger reduction of oscillation amplitude with depth, so that the waves are more concentrated near the surface. Generally, it is seen that 
the modes are confined near the surface layer of $\sim$ 15 Mm. The dispersion equation is similar to that of equatorially trapped waves with $\varepsilon \gg 1$. The observable frequencies of the Rossby waves with $m=1$ and $m=2$ are in the range of $0.65 \, \Omega < \sigma_{obs}<\Omega$ and  $1.4 \, \Omega < \sigma_{obs}<2 \Omega$, respectively, depending on stellar rotation, radius and surface temperature (see Figure 4). Inertia-gravity and Rossby-gravity modes may have the observable frequency in the interval of $5 \, \Omega < \sigma_{obs}<20 \, \Omega$.

Non-spherical distortion of rotating star may influence the frequencies and the spatial structures of Rossby waves. Let us estimate the influence of non-sphericity  from the analysis of Provost et al. (1981). They expanded all physical quantities in small parameter $(\Omega/\Omega_g)^2$, where $\Omega_g=\sqrt{G M/R^3}$ is the characteristic frequency of the star. Consequently, they assumed the frequency of Rossby waves as $\sigma=\sigma_0(1+(\Omega/\Omega_g)^2\sigma_1)$, where $\sigma_0$ is the frequency of 2D classical Rossby waves and $\sigma_1$ the first order frequency. Therefore, the correction due to the 3D consideration and non-spherical distortion is $(\Omega/\Omega_g)^2\sigma_1$. The value of $(\Omega/\Omega_g)^2$ is $\approx 1.7\cdot10^{-5}$ for a star with solar radius, mass and angular frequency, while it is $\approx 1.7\cdot10^{-3}$ for a star with 10 times faster rotation (i.e. with the period of 2.6 days). On the other hand, largest frequency correction for radiative stars according to Provost et al. (1981) is $\sigma_1=-1.121$, which corresponds to the mode with $n=3$, $l=3$, $m=1$ and $k=4$, where $n$ is the polytropic index, $l$ and $m$ are the latitudinal and azimuthal numbers, and $k$ is the number of radial nodes (see the fifth raw of the first column in the table 1 of Provost et al. (1981)). Consequently,
typical maximal correction of 2D Rossby wave frequency due to the non-spherical distortion is $2\cdot10^{-5}$ for a slowly rotating star like the Sun, and $2\cdot10^{-3}$ for a rapidly rotating star with the period of 2.6 days. Hence, the correction due to the non-spherical distortion is negligible for most cases. The situation can be changed for very rapidly rotating stars with period of $<0.1$ days. Therefore, our analysis may not be valid for these extreme cases.

It is known that the magnetic field has significant influence on the dynamics of Rossby waves \citep{Zaqarashvili2007, Marquez2017, Zaqarashvili2018, Dikpati2018, Dikpati2020, Zaqarashvili2021}. Therefore, the observed frequency of Rossby waves and their temporal variations  might be used for seismic estimations of the magnetic field strength near the surface and in interiors of stars \citep{Gurgenashvili2016,Zaqarashvili2021,Gurgenashvili2022}. But more study (observational and theoretical) is surely required to this direction in the future.    

To obtain the dispersion relations and vertical structure of Rossby waves we considered the uniform rotation of stars. On the other hand, latitudinal differential rotation may lead to the large-scale instabilities on stars \citep{Watson1981,Gilman1997,Gilman2007}, which may also lead to the instability and frequency modification of Rossby waves \citep{Zaqarashvili2010,Gizon2020,Gizon2021}. Therefore, the inclusion of the latitudinal differential rotation in the consideration is desired in the future.  

One can argue that anti-symmetric modes (with odd $l$) have small contributions in stellar light curves as the southern and northern hemispheric parts of radiance may balance each other. This effect is unimportant for the stars with the rotation axis being nearly parallel to the line of sight. Therefore, in most stars one can expect to observe only the symmetric modes with even $l$.

 As it was discussed above, the rate of temperature gradient, $\epsilon$, determines the equivalent depth, $h$, and hence defines the vertical structure and frequency of modes. On the other hand, systematic observations of wave frequency may lead to the estimation $\varepsilon$. Then one can determine the equivalent depth of corresponding mode and hence roughly estimate the vertical temperature gradient in stellar interiors. This might be an useful tool for stellar seismology. 

Considering the equatorial beta-plane approximation and linear vertical gradient of temperature in the interior of radiative stars led to the exact analytical solutions for Rossby and inertia-gravity waves. Oscillation spectra and radial structure of the waves with different wavenumbers are obtained. The waves may affect the light curves of stars, therefore they could be further observed by recent space missions. The observed Rossby waves may be used for seismology of stars with different spectral classes being at different evolutionary phases.   
\\
\\
\\
\textit{Acknowledgements}: MA was supported by Volkswagen Foundation and by Shota Rustaveli National Science Foundation Georgia (SRNSFG) [Grant number N04/46-9] in the framework of the project "Structured Education in Quality Assurance Freedom to Think". TVZ was supported by the Austrian Fonds zur F{\"o}rderung der Wissenschaftlichen Forschung (FWF) project P30695-N27. This paper resulted from discussions at workshops of ISSI (International Space Science Institute) team (ID 389) “Rossby waves in astrophysics” organised in Bern (Switzerland).


\begin{thebibliography}{}

\bibitem[Chelton and Schlax (1996)]{Chelton1996} Chelton, D. B., Schlax, M. G., 1996, Science, 1996, 272, 5259, 234-238

\bibitem[Damiani et al.(2020)]{Damiani2020} Damiani, C., Cameron, R. H., Birch, A.C., Gizon, L., 2020, A \& A, 637, A65

\bibitem[Dikpati et al.(2018)]{Dikpati2018} Dikpati, M., McIntosh, S. W., Bothun, G., Cally, P. S., Ghosh, S. S., Gilman, P. A., Umurhan, O. M., 2018, \apj, 853, 144

\bibitem[Dikpati et al. (2020)]{Dikpati2020} Dikpati, M., Gilman, P. A., Chatterjee, S., McIntosh, S. W., Zaqarashvili, T. V., 2020, \apj, 896, 141

\bibitem[Eliasen and Machenhauer (1965)]{Eliasen1965} Eliasen, E., Machenhauer, B., 1965, Tellus, 17, 2, 220

\bibitem[Gill (1982)]{Gill1982} Gill, A. E., 1982, Academic Press, London

\bibitem[Gilman and Fox (1997)]{Gilman1997}Gilman, P. A. and Fox, P. A., 1997, \apj, 1997, 484, 439

\bibitem[Gilman et al.(1997)]{Gilman2007}Gilman, P. A., Dikpati, M., Miesch, M. S., 2007, \apjs, 170, 203

\bibitem[Gizon et al.(2021)]{Gizon2021}Gizon, L., Cameron, R. H., Bekki, Y., Birch, A. C., Bogart, R. S., Brun, A. S., Damiani, C., Fournier, D., Hyest, L., Jain, K., Lekshmi, B., Liang, Z.-C., Proxauf, B., 2021, A \& A, 652, L6

\bibitem[Gizon et al.(2020)]{Gizon2020}Gizon, L., Fournier, D., Albekioni, M., 2020, A\&A, 642, A178

\bibitem[Gurgenashvili et al.(2016)]{Gurgenashvili2016}Gurgenashvili, E., Zaqarashvili, T. V., Kukhianidze, V., Oliver, R., Ballester, J. L., Ramishvili, G., Shergelashvili, B., Hanslmeier, A., and Poedts, S., 2016, \apj, 826, 55

\bibitem[Gurgenashvili et al.(2022)]{Gurgenashvili2022}Gurgenashvili, E., Zaqarashvili, T. V., Kukhianidze, V., Reiners, A., Reinhold, T., Lanza, A. F., 2022, \aa, 660, A33

\bibitem[Hadley(1735)]{Hadley1735} Hadley G., 1735, Philos. Trans. R. Soc. Lond. 34, 58

\bibitem[Hanasoge and Mandal(2019)]{Hanasoge2019}Hanasoge, S., Mandal, K., 2019, \apjl, 871, 2, L32

\bibitem[Hanson et al. (2022)]{Hanson2022} Hanson, C. S., Hansagone, S., Sreenivasan, K. R., 2022, Nature Astronomy, Online publication 

\bibitem[Haurwitz (1940)]{Haurwitz1940} Haurwitz, B., 1940, Journal of Marine Research, 3, 254

\bibitem[Henneco (2021)]{Henneco2021}Henneco, J., Van Reeth, T., Prat, V., Mathis, S., Mombarg, J. S. G., Aerts, C., 2021, A \& A, 648, A97

\bibitem[Hill et al. (2000)]{Hill2000} Hill, K. L., Robinson, I. S., Cipollini, P., 2000, Journal of Geographical Research, 105, C9, 21927

\bibitem[Hirooka and Hirota(1989)]{Hirooka1989} Hiroota, T., Hirota, I., 1989, Pure and Applied Geophysics, 130, 277 

\bibitem[Hough(1897)]{Hough1897} Hough, S. S., 1897, Philosophical Transactions of the Royal Society of London. Series A, 189,  201

\bibitem[Hough(1898)]{Hough1898} Hough, S. S., 1898, Philosophical Transactions of the Royal Society of London. Series A, 191, 139

\bibitem[Hövmoller (1949)]{Hovmoller1949} Hovermoller, E., 1949, Tellus, 1, 2, 62

\bibitem[Jeffery (2020)]{Jeffery2020} Jeffery, C.S., 2020, Monthly Notices of the Royal Astronomical Society, 496, 1, 718

\bibitem[Krista et al. (2017)]{Krista2017} Krista, L. D., Reinard, A. A., 2017, \apj, 839, 50

\bibitem[Lanza et al. (2019)]{Lanza2019} Lanza, A. F., Gizon, L., Zaqarashvili, t. V., Liang, Z. -C., Rodenbeck, K., 2019, A \& A, 623, A50

\bibitem[Laplace (1893)]{Laplace1893} Laplace, P. S., 1893, Oeuvres 9, 71

\bibitem[Li et al.(2019)]{Li2019} Li, G., Van Reeth, T., Bedding, T. R., Murphy, S. J., Antoci, V., 2019, in Monthly Notices of the Royal Astronomical Society, 487, 1, 782

\bibitem[Liang et al.(2019)]{Liang2019} Liang, Z. -C., Gizon, L., Birch, A. C., Duvall, T. L., 2019, A \& A, 626, 3

\bibitem[Lindzen (1967)]{Lindzen1967} Lindzen, R.S., 1967, in Monthly Weather Review, 95, 7, 441

\bibitem[Lindzen et al.(1984)]{Lindzen1984} Lindzen, R. S., Straus, D. M., Katz, B., 1984, Journal of the Atmospheric Science, 41, 8, 1320

\bibitem[Longuet-Higgins (1968)]{Longuet-Higgins1968}Longuet-Higgins, M. S., 1968, 
    Philosophical Transactions for the Royal Society of London. Series A, 262, 511

\bibitem[Löpten et al. (2018)]{Lopten2018} Löpten, B., Gizon, L., Birch, A. C., Schou, J., Proxauf, B., Duvall, T. L., Bogart, R. S., Christensen, U. R., 2018, Nature Astronomy, 2, 568

\bibitem[Madden (2007)]{Madden2007} Madden, R. A., 2007, Tellus, A59, 5, 571

\bibitem[M{\'a}rquez-Artavia et al. (2017)]{Marquez2017} 
M{\'a}rquez-Artavia, X., Jones, C. A., Tobias, S.M., Rotating magnetic shallow water waves and instabilities in a sphere, Geophys. Astrophys. Fluid Dyn.,2017, 111, 282

\bibitem[Matsuno(1966)]{Matsuno1966} Matsuno, T., 1966, Journal of the Meteorological Society of Japan. II, 44, 1, 25

\bibitem[McIntosh et al.(2017)]{McIntosh2017} McIntosh, S. W., Cramer, W. J., Marcano, M. P., Leamon, R. J., 2017, Nature Astronomy, 1, 0086 

\bibitem[Papaloizou and Pringle(1978)]{Papaloizou1978}Papaloizou J., Pringle J. E., 1978, MNRAS, 182, 423

\bibitem[Pedlosky(1987)]{Pedlosky1987} Pedlosky, J., 1987, Springer

\bibitem[Platzman(1968)]{Platzman1968} Platzman, G. W., Royal Meteorological Society, 94, 401, 225

\bibitem[Provost et al.(1981)]{Provost1981} Provost, J., Barthomieu, G., Rocca, A., 1981, A \& A, 94, 126

\bibitem[Proxauf et al.(2020)]{Proxauf2020} Proxauf, B., Gizon, L., Lopten, B., Schou, J., Birch, A. C., Bogart, R. S., 2020, A \& A, 634, A44, 16

\bibitem[Rossby(1939)]{Rossby1939} Rossby, C. -G., 1939, Journal of  Mar. Res. 2, 38

\bibitem[Saio(1982)]{Saio1982} Saio, H., 1982, Astrophysical Journal, Part 1, 256, 717

\bibitem[Saio et al.(2018)]{Saio2018} Saio, H.,  Kurtz, D. W., Murphy, S. J., Antoci, V. L. Lee, U., 2018, Monthly Notices of the Royal Astronomical Society, 474, 2, 2774

\bibitem[Saio and Kurtz(2022)]{Saio2022} Saio, H., Kurtz, D.W., 2022, in Monthly Notices of the Royal Astronomical Society, 511, 1, 560

\bibitem[Salby(1984)]{Salby1984} Salby, M. L., 1984, Reviews of Geophysics, 22, 2, 209

\bibitem[Samadi-Ghadim et al.(2020)]{Samadi2020} Samadi-Ghadim, A., Lampens, P., JAssur, D. M., Jofre, P., 2020, A \& A, 638, A57, 19

\bibitem[Takata et al.(2020)]{Takata2020} Takata, M., Ouazzani, R. -M., Saio, H., Ballot, J., Antoci, V., Salmon, S. J. A. J., 2020, A \& A, 644, A138, 12

\bibitem[Taylor(1936)]{Taylor1936} Taylor, G.I., 1936, The Royal Society Publishing, 156, 888

\bibitem[Van Reeth et al.(2016)]{VanReeth2016} Van Reeth, T., Tkachenko, A., Aerts, C., 2016, A \& A, 593, A120, 13

\bibitem[Yanai and Lu(1983)]{Yanai1983} Yanai, M., Lu, M. M., 1983, Journal of the Atmospheric Science, 40, 12, 2785

\bibitem[Watson(1981)]{Watson1981}Watson, M., 1981, Geophysical and Astrophysical Fluid Dynamics, 16, 285

\bibitem[Zaqarashvili et al.(2007)]{Zaqarashvili2007} Zaqarashvili, T. V., Oliver, R., Ballester, J. L., and Shergeashvili, B. M, 2007, A \& A, 470, 815

\bibitem[Zaqarashvili et al.(2010)]{Zaqarashvili2010} Zaqarashvili, T. V., Carbonell, M., Oliver, R., Ballester, J. L., 2010, \apj, 709, 2, 749

\bibitem[Zaqarashvili(2018)]{Zaqarashvili2018} Zaqarashvili, T. V., 2018, \apj, 856, 1, 32, 11

\bibitem[Zaqarashvili et al.(2021)]{Zaqarashvili2021} Zaqarashvili, T.V., Albekioni, M., Ballester, J. L., Bekki, Y., Biancofiore, L., Birch, A. C., Dikpati, M., Gizon, L., Gurgenashvili, E., Heifetz, E., Lanza, A. F., McIntosh, S. W., Ofman, L., Oliver, R., Proxauf, B., Umurhan, O. M., Yellin-Bregovoy, R., 2021, Space Science Review, 217, 15


\end{thebibliography}
\end{document}